\documentclass[aps,prb,superscriptaddress,twocolumn]{revtex4-1}
\usepackage{color}

\usepackage{amsmath,amsfonts,amsthm}
\usepackage{chngcntr}
\usepackage{gensymb}
\usepackage{amssymb}
\usepackage{soul}
\usepackage[T1]{fontenc} % Use 8-bit encoding that has 256 glyphs
\usepackage[english]{babel} % English language/hyphenation
\usepackage{multirow}
\usepackage{graphicx} % For including figures
\usepackage{lmodern}
\usepackage{mathtools}
\usepackage{hyperref}   %\url{http://example.com/}
\usepackage{lipsum} % Used for inserting dummy 'Lorem ipsum' text into the template
\usepackage{fancyhdr} % Custom headers and footers
\numberwithin{equation}{section} % Number equations within sections (i.e. 1.1, 1.2, 2.1, 2.2 instead of 1, 2, 3, 4)
\numberwithin{figure}{section} % Number figures within sections (i.e. 1.1, 1.2, 2.1, 2.2 instead of 1, 2, 3, 4)
\numberwithin{table}{section} % Number tables within sections (i.e. 1.1, 1.2, 2.1, 2.2 instead of 1, 2, 3, 4)
\counterwithout{figure}{section}
\counterwithout{table}{section}
\begin{document}
\newcommand{\hbcomm}[1]{\textcolor{blue}{\textit{\textbf{HB:} #1}}}
\newcommand{\hb}[1]{\textcolor{blue}{#1}}

\title{Charge Order Driven Multiferroic Behaviour in Sr$_4$Fe$_6$O$_{12}$: An \textit{Ab-initio} Study} 
\author{ Arindam Sarkar}\email{arindam@phy.iitb.ac.in}
\affiliation{Department of Physics, Indian Institute of Technology, Bombay, Powai, Mumbai 400076, India}  
\author{Hena Das}
\affiliation{ Laboratory for Materials and Structures, Institute of Integrated Research (IIR), Institute of Science Tokyo, 4259 Nagatsuta, Midori-ku, Yokohama 226-8503, Japan.}
\author{Prashant Singh}
\affiliation{Ames Laboratory, U.S. Department of Energy, Iowa State University, Ames, Iowa 50011-3020, USA}
\author{Aftab Alam}\email{aftab@iitb.ac.in}
\affiliation{Department of Physics, Indian Institute of Technology, Bombay, Powai, Mumbai 400076, India} 

%\date{\today}

\begin{abstract}
In this letter, we report the structural, electronic and ferroelectric properties of the layered mixed-valent transition-metal compound, Sr$_{4}$Fe$_{6}$O$_{12}$ (SFO). We demonstrate how SFO undergoes a phase transition from a high-temperature (T) centrosymmetric tetragonal phase ($P4_{2}/mnm$) to a low-T polar orthorhombic phase ($Pmn2_{1}$). The transition is primarily driven by charge ordering at tetrahedral Fe-layer creating Fe$^{3+}$ and Fe$^{2+}$ cations between two edge sharing tetrahedra. This charge ordering induces electronic polarization, which is remarkably larger (3.5 times) in magnitude than ionic polarization and oppositely directed, giving a net polarization of 0.05 C/m$^2$ which is comparable to the state-of-the-art rare-earth nickelets and manganite perovskites. The direction of structural distortion, governed by the polar mode irrep $\Gamma_{5}^{-}$, depends sensitively on the type of magnetic ordering in the Fe-octahedral layer. Consequently, both the ionic and electronic polarization directions are influenced by magnetic ordering, suggesting the potential for multiferroic behavior with strong magneto-electric coupling in this material.
\end{abstract}

\maketitle

Ferroelectric (FE) materials have attracted tremendous scientific interest due to their diverse technological applications, such as those in memory devices, filters, and high-performance insulators \cite{lines2001principles}. Recent studies have focused on multiferroic oxides, which exhibit both ferroelectricity and magnetic ordering \cite{schmid2001materials,wang2009multiferroicity}, enabling coupled electric and magnetic properties \cite{spaldin2005renaissance, cheong2007multiferroics}. Such coupling offers unique opportunities for manipulating one property via the other, expanding their functional applications \cite{pokhrel2023ferroelectric,fiebig2005revival}. In particular, spin-frustrated, charge-ordered systems stand out due to their strong spin-charge coupling, where ferroelectricity arises from alternating short-and-long-bond breaking inversion symmetry \cite{van2008multiferroicity,ikeda2005ferroelectricity,de2012charge,radaelli2002formation, giovannetti2009multiferroicity1, tokura2000orbital}. Although several multiferroic transition metal oxides (TMOs) have been reported to date, studies on ferroelectricity in spin-frustrated systems are extremely limited. The strong spin-charge coupling and magnetic properties of mixed/fractional valence TMOs make them an ideal candidate for advanced multifunctional devices.

In this letter, we present an interesting TMO, i.e., Sr$_4$Fe$_6$O$_{12}$ (SFO), that shows charge ordering and geometric spin frustration, which makes it an unconventional multiferroic candidate. Unlike typical multiferroics, where ionic and electronic polarizations align, the SFO exhibits an unusual antiparallel alignment of these components, with electronic polarization dominating over the ionic contribution. This behavior contrasts with systems like HoMn$_2$O$_5$(RMn$_2$O$_5$, R= Ho,Tb,Y,Eu,etc.) \cite{giovannetti2008electronic}, where equal but opposite ionic and electronic polarizations cancel out. The only other known material showing such antiparallel polarization is the quasi-one-dimensional organic crystal TTF-CA \cite{giovannetti2009multiferroicity,kobayashi2012electronic}. The dominant nature of electronic polarization makes SFO a unique inorganic compound that provides a valuable platform to study unconventional polarization mechanisms in multiferroics.

To the best of our knowledge, SFO is the first TMO in which electronic polarization is almost 3.5 times higher than the ionic component. This makes total polarization  of the order of 0.05~C/m$^2$, which is comparable to the state-of-the-art rare-earth nickelets and manganite perovskites\cite{giovannetti2009multiferroicity1,picozzi2007dual}. The SFO shows an insulating behavior both in charge ordered and disordered phases. Ferroelectricity kicks in when the compound goes through a structural phase transition from non-polar to polar phase, producing charge ordering and lifting the inversion symmetry. Our simulation reveals that the atomic displacements accompanying this structural distortion are very sensitive to the collinear magnetic configurations at octahedral Fe-sublattices. As a result, the direction of polarization is also intricately linked to the specific magnetic ordering, underscoring the strong interplay among structure, electronic, and magnetic degrees of freedom.

\begin{figure*}[th]
\centering
\includegraphics[scale=0.22]{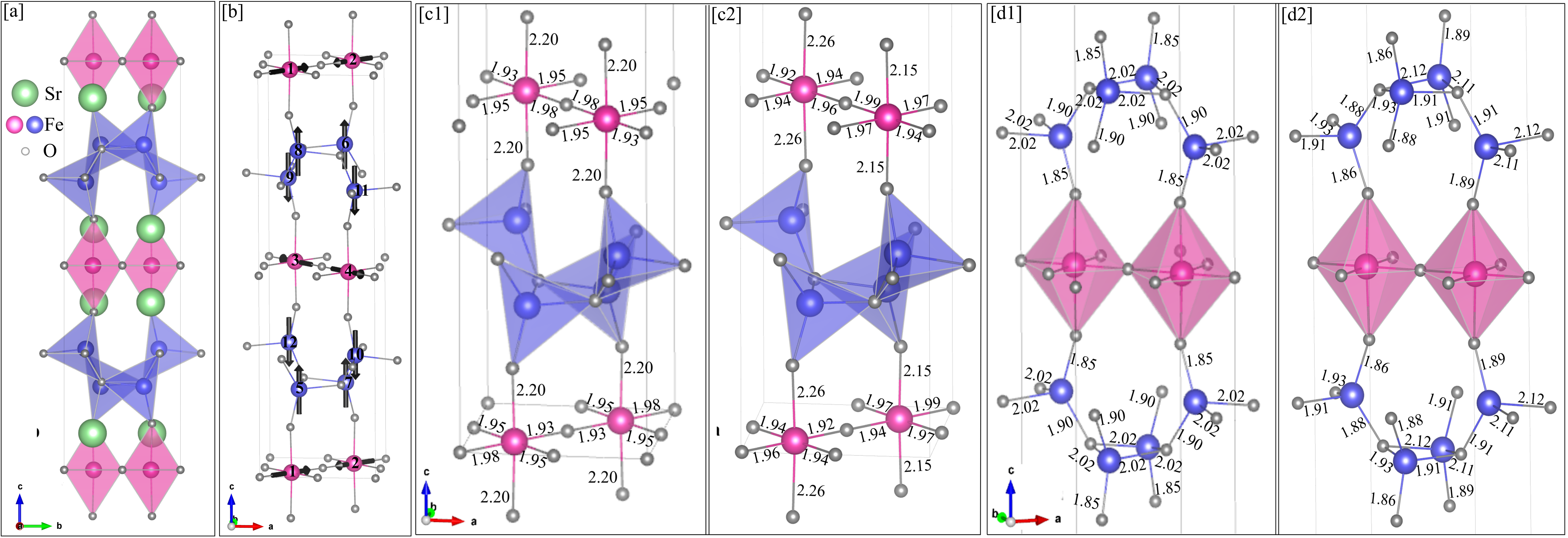}
\caption{[a] Crystal structure of Sr\(_{2}\)Fe\(_{6}\)O\(_{12}\) at 2 K (unrelaxed). [b] Experimentally observed non-collinear magnetic ordering, henceforth termed as NC-AFM. Fe atoms are numbered as per the experimentally provided structure\cite{lu2013sr4fe6o12}. Octahedrally coordinated Fe-O bonds in the [c1] unrelaxed and [c2] relaxed structure with NC-AFM ordering. Fe-O bond lengths in tetrahedral coordination for the [d1] unrelaxed and [d2] relaxed structure with NC-AFM ordering.}
\label{str}
\end{figure*}

SFO was predicted long ago in connection with the Ruddlesden-Popper phase derived compound Sr$_{4}$Fe$_{6}$O$_{13-\delta}$ \cite{rossell2005transmission,mellenne2004oxygen}. Initially, it was proposed to have an orthorhombic phase with space group (sg) $Fmmm$, consisting of octahedral layers (FeO$_{6}$) separated by two tetragonal pyramidal layers (FeO$_{5}$). However, recent experiments show that the SFO crystallizes in a tetragonal phase (sg: $P4_{2}/mnm$) \cite{lu2013sr4fe6o12}, where the octahedral layers are separated by two connected tetrahedral layers. The octahedra within each layer are connected via corner-sharing oxygen, while the tetrahedra share edges and connect with adjacent layers through corner-sharing oxygen (see Fig.~\ref{str}[a]). Edge-sharing Fe tetrahedra in adjacent layers align along mutually $\perp$ [110] and [1$\bar{1}$0] crystallographic directions. In this centrosymmetric tetragonal phase, the octahedral and tetrahedral Fe occupy the 4$f$ and 8$j$ Wyckoff sites, respectively, surrounded by 5 inequivalent oxygens (O$_{1}$ to O$_{5}$)\cite{footnote1}. The octahedral Fe are in the +3 oxidation state, leading to charge ordering between edge-sharing tetrahedra, forming a dimer-like pattern, resulting in a 2:1 ratio of Fe$^{3+}$:Fe$^{2+}$ in the charge-ordered state.

\textit{Ab-initio} calculations were performed using density functional theory (DFT),\cite{hohenberg1964inhomogeneous} as implemented in Vienna \textit{ab initio} simulation package (VASP)\cite{kresse1996efficiency, kresse1993ab}. To capture the strong Fe-$d$ correlation, we applied Hubbard parameters according to the Dudarevs' scheme \cite{anisimov1997first,dudarev1998electron} with values, U=4.5 eV, J=0.9 eV. These values (U,J) are similar to the one used in literature for charge ordered Fe$_3$O$_4$ compound \cite{jeng2004charge}. The U values (0.0 to 10.0 eV) were carefully optimized to understand the correlation between structure and ferroelectric polarization, while the J was fixed to 0.9 eV. Further computational details are provided in the supplementary material (SM)\cite{supp}; see also \cite{blochl1994projector,perdew1996generalized}.

Figure \ref{str}[b] shows the experimentally reported magnetic structure, i.e., a non-collinear (NC) spin-arrangement in the octahedral layer (referred to as NC-AFM), where the moments within each octahedral layer are antiparallel in the crystallographic \(ab\)-plane along the Fe-O-Fe bond direction. As a result, the moments in two consecutive octahedral layers are oriented at 90$^{\circ}$ angle. Furthermore, the moments within each octahedral layer show weak ferromagnetic (FM) alignment along the \(z\)-axis, which is antiparallel to the connected tetrahedral layers. This behavior arises out of magnetic frustration, stemming from a unique combination of valency and structural geometry that leads to competing FM and AFM super-exchange interactions between the octahedral and tetrahedral layers. In addition to the NC spin arrangement, four other energetically competing collinear spin configurations, namely FM, A-AFM, B-AFM, and C-AFM are considered in the present study (see Fig.~S1 and Table~S4 of SM\cite{supp}). Among these, only the A-AFM and B-AFM reproduce the experimentally observed electronic behavior. Therefore, in this work, we focus on the microscopic understanding of NC-AFM along with the A-AFM and B-AFM orderings, and reveal the competing electronic effects.

In the unrelaxed structure, the Fe-O bond lengths in all octahedra are identical (see Fig.~ \ref{str}[c1]). However, after {\it ionic} relaxation of the NC-AFM structure, the planar bond lengths of two corner-sharing octahedra alternately decrease and increase (see Fig. ~\ref{str}[c2]). The axial bonds exhibit an opposite trend. Consequently, the two octahedra become inequivalent, with their respective octahedral volumes changing to 11.06~\AA$^3$ and 11.33~\AA$^3$, compared to the uniform unrelaxed value of 11.18~\AA$^3$. Similarly, in the unrelaxed structure, all tetrahedra have equal Fe-O bond lengths (see Fig. ~\ref{str}[d1]). After relaxation, these bond lengths alternately decrease and increase between two edge-sharing tetrahedra within each layer (see Fig. ~\ref{str}[d2]), resulting in a breathing-like distortion. This in-equivalence is reflected in the tetrahedral volumes, which change to 3.40~\AA$^3$ and 3.84~\AA$^3$ from their uniform unrelaxed value of 3.56~\AA$^3$. This change in volume is attributed to reduced planar bond lengths that shortens the distance between two tetrahedral Fe, i.e., from 2.90~\AA\ (unrelaxed) to 2.84~\AA\ (relaxed), consistent with experiments \cite{lu2013sr4fe6o12}. The planar Fe-O-Fe bond angles decrease from 92$^\circ$ to 89$^\circ$. Additionally, not only do the Fe-O bonds between two edge-sharing tetrahedra differ from each other, but the two planar bonds within each tetrahedron also become unequal. This breaks the inversion symmetry, thereby reducing the crystallographic symmetry. For collinear magnetic ordering, however, the distortion of octahedra and their connected tetrahedra depends on the specific magnetic ordering in the octahedral layer (for more details, see Sec. V and VI of SM\cite{supp})

It should be noted that after {\it full} (ionic+volume/shape) relaxation, a similar trend in structural reconstruction is observed. However, the planar Fe-O bond lengths turn out to be larger than those in the ionic-relaxed structure. More details can be found in footnote \cite{footnote2}. The dependence of lattice constants on Hubbard U for NC-AFM ordering shows that the length of the \(c\)-axis increases, while the lengths of \(a\)- and \(b\)-axes decrease compared to the experimentally reported values when \(U > 4.5\)~eV (see Fig.~S2). For \(U < 4.5\)~eV, the system remains metallic with no charge ordering. These findings validate our choice of \(U = 4.5\)~eV. Similar results are obtained for collinear orderings (A- and B-AFM) as well (see Sec. V of SM\cite{supp}).

The direction of polarization depends on the orientation of individual bond distortions, which are influenced by the magnetic structure. 
In all of cases, we observed that while the magnitude of polarization remains consistent, its direction varies within the crystallographic \(ab\)-plane due to variations in the orientations of structural distortions.
The present theoretical findings can be extremely useful in identifying the type of magnetic ordering through the measured polarization direction or for stabilizing specific magnetic ordering by cooling the material under an applied electric field. 

\begin{figure}[t]
\centering
\includegraphics[scale=0.4]{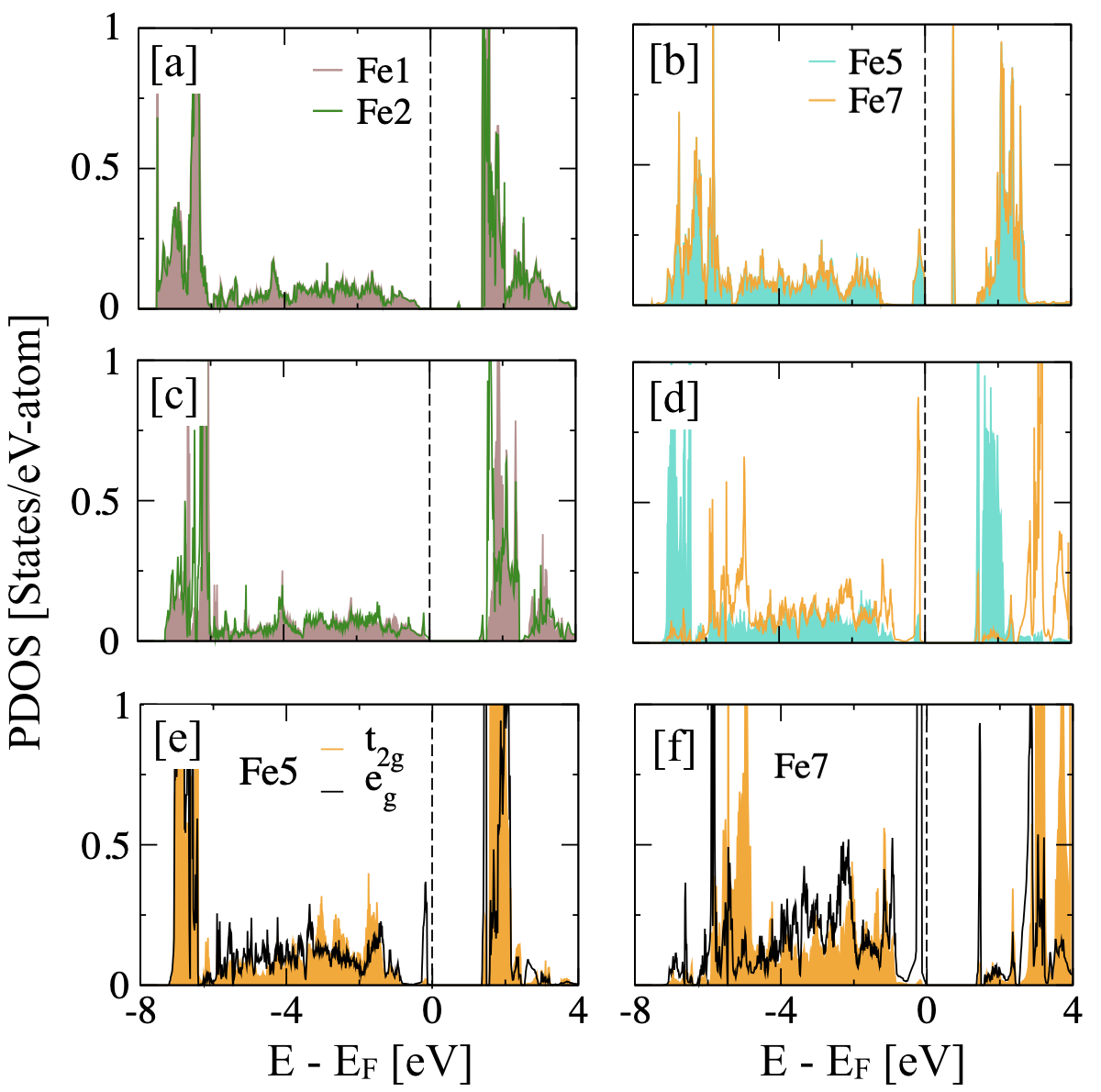}
\caption{For NC-AFM ordering of SFO, atom projected density of states for [a] corner-sharing octahedral (Fe1, Fe2) and [b] edge-sharing tetrahedral (Fe5, Fe7) Fe atoms in the experimental structure. [c,d] Same as Fig. [a,b] but for theoretically relaxed structure. [e,f] partial t$_{2g}$ and e$_g$ density of states of Fe on edge-sharing tetrahedra in the relaxed structure. PDOS of only one spin channel is shown as the other is mirror symmetric.}
\label{pes_dos}
\end{figure}

The in-equivalency of the polyhedra is mainly driven by the change in the positions of the tetrahedral Fe and oxygen at the 8$j$ Wyckoff sites in the tetragonal structure (see Table-S4 in SM \cite{supp}). As a result, the overall symmetry of the structure reduces from tetragonal ($P4_{2}/mnm$) to orthorhombic ($Pmn2_{1}$) phase under ionic relaxation. On the other hand, upon full relaxation, the tetragonal symmetry changes to monoclinic ($Pm$(6)), both remains noncentrosymmetric polar structures. This indicates that the reported experimental structure may be metastable at low temperatures (T) and could undergo a structural phase transition to a lower symmetry structure. Such a structural transition can be analyzed via the group-subgroup relation between the parent and distorted structures \cite{stokes2016isosubgroup} (see Sec. II of SM\cite{supp}). A brief description of various distortions and the corresponding order parameters (OP) of ionic-relaxed structure can be found in footnote\cite{footnote3} (also see Sec. IV and V of SM\cite{supp}).

Figures~\ref{pes_dos}[a],[b] show the atom-projected density of states (DOS) of corner-sharing octahedral (Fe1, Fe2) and edge-sharing tetrahedral (Fe5, Fe7) in the experimentally reported centrosymmetric structure (P4$_2$/mnm) with NC-AFM ordering. Because of the mirror-symmetric nature of up/down-spin DOS in AFM phase, we only show the up-spin channel. The Fe DOS clearly shows that both iron octahedra and tetrahedra are equivalent. The atom projected Fe DOS both in octahedral and tetrahedral setting for the relaxed structure are shown in Figs. ~\ref{pes_dos}[c],[d]. Notably, the relaxed structure shows the appearance of in-equivalency between octahedral and tetrahedral Fe. This leads to change in the oxidation state of the tetrahedral Fe, indicating the emergence of charge ordering that was absent in the unrelaxed structure. The excess electron goes to the e$_g$ orbital (more precisely to the d$_{x^2-y^2}$ orbital, as evident from the partial DOS plots in Figs. ~\ref{pes_dos}[e],[f]). For the collinear magnetic arrangements (see Figs. ~S9 \& S10 in SM\cite{supp}), similar behavior is observed, where the fractional valency of tetrahedral Fe in the unrelaxed centrosymmetric structure (Fe$^{2.5+}$) is lifted via a structural phase transition, creating Fe$^{3+}$ and Fe$^{2+}$ valency in each layer between two edge-sharing tetrahedra. Both the unrelaxed and relaxed structures are insulators. With collinear magnetic ordering, the band gap (E$_g$) is 0.74 eV for the unrelaxed structure and 1.4 eV for the relaxed structure. With NC-type AFM ordering, E$_g$ is 0.73 eV and 1.34 eV for the unrelaxed and relaxed structures, respectively. Further comparison of the band structure for collinear ordering between the relaxed and unrelaxed structures, including charge transfer and bonding features, is shown in Sec. VII of SM\cite{supp}.

\begin{figure*}[t]
\centering
\includegraphics[scale=0.55]{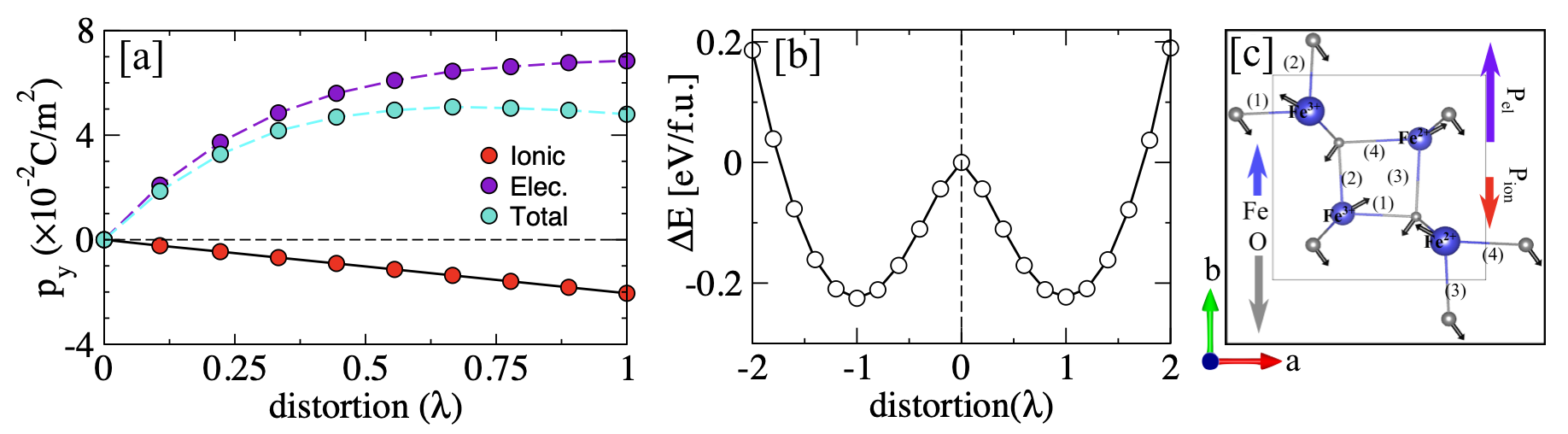}
\caption{For NC-AFM ordering of SFO,
[a] electronic and ionic component of polarization (along the b-axis) vs. ferroelectric distortion (\(\lambda\)). [b] Relative energy (\(\Delta\)E) vs. \(\lambda\). [c] Schematic of displacements on the Fe-O tetrahedral plane  with respect to their positions in the tetragonal phase. Smaller and larger atoms are in successive layers along the c-axis. Bond labels (1) to (4) represent bond lengths in ascending order: 1.91, 1.93, 2.11, and 2.12~\AA. Net displacements of Fe and O are shown along with the direction of polarization components.  }
\label{pol}
\end{figure*}

Figure~\ref{pol}[b] shows the variation of relative energy (\(\Delta\)E) with respect to distortion amplitude (\(\lambda\)) due to the irrep \(\Gamma^{-}_5\) for NC-AFM ordering. The double-well-like shape of the energy surface with a negative curvature at the origin clearly indicates the instability of the centrosymmetric structure at low T, leading to a spontaneous symmetry lowering from the tetragonal to the orthorhombic phase. The energy surface is nearly a perfect double-well (the energy difference is about 2.1~meV/formula unit (fu.)), indicating the close degeneracy of the positive and negative distortions, or FE-switching, where either one of the octahedra and the corresponding connected tetrahedra in each layer can be dilated or contracted. However, the energy surface is not a perfectly symmetric double-well for two collinear AFM configuration (see Fig.~S5 in SM \cite{supp}). This indicates that, with the same collinear ordering, structures with positive and negative distortions are not symmetric. For negative distortion, the magnetic alignment at the octahedral layers needs to be reversed with respect to the ordering used for relaxation (Fe$^{oct}_\uparrow$-Fe$^{oct}_\downarrow$ \(\rightarrow\) Fe$^{oct}_\downarrow$-Fe$^{oct}_\uparrow$). In other words, the direction of distortion and the direction of polarization can be reversed by reversing the collinear magnetic ordering for A- and B-AFM orderings. Similar behavior is also reported in the multiferroic perovskite HoMnO$_3$ \cite{picozzi2007dual}. The energy difference between A-AFM (B-AFM) and its spin-reversed counterpart A$^\prime$-AFM (B$^\prime$-AFM) orderings with negative distortion is 34 meV/fu. These results indicate the presence of strong spin-charge-lattice coupling and possible magnetic anisotropy. The total energy of A- and B-AFM ordering is very close to each other, with the former being 0.11 meV/fu. higher in energy. Meanwhile, the structure relaxed with NC-type ordering is 0.426 eV/fu. lower in energy compared to the B-AFM ordering. The depth of the energy well for both A- and B-AFM is $\sim$0.242 eV/fu., whereas for NC-type AFM ordering, it is $\sim$0.225 eV/fu.

Next, we evaluated the electronic contributions to polarization using the Berry phase method\cite{king1993theory,resta1994macroscopic} as implemented in VASP\cite{nunes2001berry,souza2002first}. Apart from a negligibly small off-diagonal component, the magnitude of polarization for all three magnetic orderings is equal, but their direction varies depending on the ordering (see Table~S8 in SM\cite{supp}). The electronic component is directed opposite to the ionic component and is 3.5 times larger in magnitude (see Fig. ~\ref{pol}[a]). For A-AFM, polarization is directed along the crystallographic a-axis (or Cartesian x-axis) whereas for NC- and B-AFM, it is directed along the crystallographic b-axis. This implies that the polarization is always in the ab-plane. This difference in polarization direction is mainly due to the magnetic interaction mediated distortion.

To better understand the above findings, we closely examined the intricate displacement pattern of tetrahedral Fe and in-plane oxygen. In the centrosymmetric structure (unrelaxed), two O$_5$ are equidistant from both tetrahedral Fe. However, due to the polar-mode distortion ($\Gamma^{-}_5$), the displacements of O$_5$ in successive layers are perpendicular to each other in the crystallographic \(ab\)-plane. The magnitude of the displacements for tetrahedral Fe (in-plane O$_5$) is the largest, i.e., 0.064~\AA\ (0.09~\AA), from their initial positions in the tetragonal phase. Displacements along the \(a\)- and \(c\)-axes compensate for each other in successive layers, whereas displacements along the \(b\)-axis add up. The net displacements of all tetrahedral Fe and O$_5$ due to the $\Gamma^{-}_5$ polar mode are along the \(b\)-axis (but in opposite directions) for NC-AFM (see Fig. \ref{pol}(c)) and B-AFM ordering (see Fig. S11 in SM\cite{supp}). In each tetrahedral layer, the magnitude of such displacement for Fe (O$_5$) is 0.03~\AA\ (0.064~\AA). Consequently, the Fe shift away from the centers of the tetrahedra, and the O$_5$ shared between edge-sharing tetrahedra become inequivalent, forming O$_5$ and O$^{\prime}_5$. This lifts the inversion symmetry. Additionally, two planar Fe-O bonds in each tetrahedron become unequal, resulting in an ionic component of polarization along the \(b\)-axis. 

A similar trend is also observed for A-AFM ordering (see Fig. S11 in SM\cite{supp}), where the net displacements of Fe and O$_5$ are of similar magnitude along the a-axis, but opposite in direction. This creates an ionic component of polarization along the a-axis. The magnitude of ionic polarization in all these cases is 0.02~C/m$^2$. The in-equivalence of O$_5$ between two edge-sharing tetrahedra and the unequal Fe-O bonds within and between each tetrahedron lead to a partial shift away from site-centered charge ordering toward bond-centered charge ordering \cite{efremov2004bond}. For a distortion that creates a longer Fe-O bond, the partial charge shift is from the cation to the anion, whereas for shorter bond, the direction of the partial charge shift is reversed \cite{ghosez1998dynamical}. We also observed that the differences in planar Fe-O bond lengths within edge-sharing tetrahedra are unequal. In the larger tetrahedron, this difference is 0.01~\AA, whereas for smaller one, it is 0.02~\AA. Consequently, the charge shift is unequal along the direction of the mismatch. For NC and B-AFM, this partial charge shift in Fe tetrahedra is along the crystallographic b-axis (see Fig. \ref{pol}[c] and Fig. S11\cite{supp}). In contrast, for A-AFM ordering, the orientation of the unequal Fe-O bonds within the tetrahedra is rotated by approximately 90$^\circ$, causing the partial charge shift to align along the a-axis, which is also the direction of electronic polarization. 

In all of these cases, the partial charge transfer occurs in the direction opposite to the anionic displacement, causing the electronic polarization to oppose the ionic component. The electronic polarization for each magnetic ordering is 0.07~C/m$^2$. It is worth noting that for both collinear arrangements, the magnetic inversion symmetry is broken even in the centrosymmetric tetragonal structure, resulting in a finite polarization (see Fig. S11 in SM\cite{supp}). Importantly, for the centrosymmetric structure, the electronic polarization vanishes when the octahedral spins are parallel in each layer (Fe$^{\text{oct}}_\uparrow$-Fe$^{\text{oct}}_\uparrow$ or Fe$^{\text{oct}}_\downarrow$-Fe$^{\text{oct}}_\downarrow$).

The Born effective charge (BEC) is an important quantity that often manifests the coupling between lattice displacements and electrostatic fields, indicating ferroelectric instability \cite{ghosez1998dynamical}. We found that the BEC tensors of all atoms, except octahedral Fe, in both structures are highly asymmetric and anisotropic in all three magnetic configurations (see Table~S9 in SM\cite{supp}). This asymmetry arises primarily from the anisotropy of the local atomic environment. in addition to the BEC tensors, the eigenvalues of the symmetric part of the tensors are also reported in Table S9. An interesting feature of the BEC tensors for tetrahedral Fe is that their \(x\)-\(y\) components are negative, which results in one diagonal element being negative and opposite in sign to their formal valency. A similar trend is observed for oxygen in the tetrahedral layer, where the \(x\)-\(y\) components and one diagonal element have positive signs. This behavior is possibly due to the dominance of electronic polarization over ionic polarization, creating effective net negative and positive charges on the cations and anions in the tetrahedral layer, respectively. Consequently, tetrahedral Fe and O effectively act as acceptor and donor sites, respectively \cite{dawber2012electrons}. For all other atoms, the signs of the tensor components are consistent with their formal valency, and their eigenvalues are greater than their valence states.

In conclusion, we report the electronic, magnetic, and multiferroic properties of Sr$_2$Fe$_6$O$_{12}$, where the electronic and ionic components of polarization are antiparallel each other. The ferroelectric (FE) behavior is driven by charge ordering between unusual edge-sharing tetrahedra with fractionally valent Fe. The structure undergoes a distortion from a non-polar tetragonal phase ($P4_{2}/mnm$) to a polar orthorhombic phase ($Pmn2_{1}$), driven by the two-dimensional irreducible representation $\Gamma^{-}_5$, which creates unequal Fe-O bonds in each tetrahedron. The electronic component of the polarization is 3.5 times larger in magnitude than the ionic component, with a net polarization of 0.05 C/m$^2$ which is comparable to the state-of-the-art rare-earth nickelets and manganite perovskites. Although the magnitude of polarization remains the same, its direction depends sensitively on the collinear or non-collinear nature of the magnetic ordering, consequently varying within the crystallographic \(a\)-\(b\) plane. To the best of our knowledge, this is the first-ever TMO reported to exhibit such exotic behavior. 

A.S. thanks IITB for providing financial assistance during this work. PS at the Ames National Laboratory was supported by the Division of Materials Science and Engineering of the Office of Basic Energy Sciences, Office of Science of the U.S. Department of Energy (D.O.E). Ames National Laboratory is operated for the U.S. DOE by Iowa State University of Science and Technology under Contract No. DE-AC02-07CH11358.


\begin{thebibliography}{56}
\bibitem{footnote1} O$_1$, O$_2$ and O$_3$ oxygen atoms in octahedral plane occupy 2$a$, 4$c$ and 2$b$ wyckoff sites respectively. O$_4$ and O$_5$ both occupy 8$j$ wyckoff sites where O$_4$ connects ocathedral and tetrahedral layers and O$_5$ are shared by tetrahedras in each tetrahedral layer.
\bibitem{footnote2} {Fe-Fe distance in tetrahedral layer is 2.87\AA. Due to larger planer Fe-O bonds, values of lattice constatnts a and b become larger as well compared to unrelaxed structure. New lattice constants, after full relaxation for A-AFM ordering, are a=5.59\AA, b=5.60\AA and c=5.44\AA. Lattice constants of fully relaxed structure with B-AFM ordering are a=5.60, b=5.59, c=19.43 and for NC- AFM ordering cell parameters are a=5.61, b=5.60, c=19.40, confirming the experimental findings that at low temperature a and b axes are elongated whereas c-axis is contracted.}
\bibitem{footnote3} {The distortion is due to zone center lattice instability described by two-dimentional irrep $\Gamma^{-}_5$. For irrep $\Gamma^{-}_5$, the order parameter (OP) can have two general components (a,b), where a,b corresponds to amplitude of distortion along specific direction in the irrep vector space. Here only one dimensional order parameter (a,0) or (0,a) is responsible for the distortion in a restricted one-dimensional space that drives the symmetry transition to orthorhombic phase. As a result wyckoff positions 4$c$, 4$f$ in tetragonal phase transforms to double 2$a$ sites and 8$j$ sites in  tetragonal phase transforms to two 4$b$ sites in the orthorhombic phase. Since $\Gamma^{-}_5$(a,0) alone is responsible for this symmetry transition, this is the primary order parameter. There are two other one dimensional secondary order parameters $\Gamma^{+}_1$ and $\Gamma^{+}_2$ (see SM\cite{supp}). Most general distortion with two dimensional order parameter $\Gamma^{-}_5$(a,b) leads to monoclinic phase ($Pm$) which is obtained after full relaxation.}
\end{thebibliography}
\end{document}